# AUTOMATIC DETERMINATION OF THE DIFFERENT CONTROL MECHANISMS IN UPRIGHT POSITION BY A WAVELET METHOD

P. Bertrand, LAPSCO UPRESA CNRS 6024, Université Blaise Pascal, Clermont-Ferrand, France
J.M. Bardet, LSP CNRS C55830, Université Paul Sabatier, Toulouse, France
M. Dabonneville, LIMOS FRE CNRS 2239, Université Blaise Pascal, Clermont-Ferrand, France
A. Mouzat, LAPSCO UPRESA CNRS 6024, Université Blaise Pascal, Clermont-Ferrand, France
P.Vaslin, LIMOS FRE CNRS 2239, Université Blaise Pascal, Clermont-Ferrand, France

*Abstract*-**A recent model to analyze the Center of Pressure trajectories is based on the fractional Brownian motion. By doing so, one note that standing still is describe by different mechanisms following the frequency. Previous studies exhibit the existence of a control mechanism which stabilize the upright position at a large enough time scales (from 0.3 s to 1.2 s depending on the method and on the authors) or equivalently at low frequencies. The different mechanisms are separated by a critical time scale or equivalently a critical frequency. This critical frequency is fundamental to understand the control mechanism of upright position : only physiological phenomenon at frequencies larger than this critical frequency could contribute to the task of maintaining equilibrium.
A new statistical method is introduced based onto the recent progress in signal processing : the wavelets analysis. The algorithm is entirely automatic. Seventeen healthy young subjects were studied under quiet-standing conditions, the mean value of critical frequency is 1.8 Hz corresponding to a mean critical time scale 0.68 s. The algorithm is entirely automatic.**
*Keywords*- **Postural Control, Stabilogram, Fractional Brownian motion, Wavelet analysis.**

## I. INTRODUCTION

A good model of control mechanism of undisturbed stance control in human upright posture is useful in applications. Several statistical studies have been proposed in the past. All are based on the center of pressure (COP), which is the resultant point of the reaction forces measured by a force platform. In the seventies, these data were analyzed as a set of points without any order, although COP is time varying and should be considered as time series or as the observation of the trajectory of a stochastic process at discrete times. In this framework the questions are :
i) which class of stochastic processes provide a good model?
ii) how determining with reliability the parameters of the process?
iii) what is the biological meaning of these parameters ?

Different models have been proposed (Ornstein-Uhlenbeck processes in [6], roughly speaking a spring with a random excitation), but the most convincing seems the model of "so-called" fractional Brownian motion (f.B.m.) introduced by Collins and De Luca. Using the stabilogram diffusion analysis, they detect two different zones on the log/log plot of the stabilogram with two different slopes : these two zones are interpreted as corresponding to two different mechanisms at high frequencies (an erratic comportment) and at low frequencies (regulation).

Moreover they determined by a graphical method a critical time corresponding to the change between the two regulation mechanisms. However the method proposed by Collins & De Luca is heuristic and depends on the subjectivity of the user, see for e.g. [7]. For studying the influence of experimental conditions on sufficiently large samples, an automatic procedure of determination of the change time is needed. A second drawback of the pioneer work of Collins & De Luca and of the use statistical method based on f.B.m is the absence of a rigorous model which forbids any justified statistical study. The aim of this paper is to propose an automatic method of detection of the change point between the different mechanisms of regulation and to apply it to a set of 17 experimental data.

## II. METHODOLOGY

*Material*

The force-plate was an AMTI model OR6-5. The signals were amplified through an AMTI model SGA6-4 amplifier. The numerical-analog converter card was a Data Translation model DT2801 (12 bits). Recorded data were the three forces and the three moments of the efforts of contact on the surface of the force-plate. From the forces and moments measured on the platform, we calculated the COP coordinates *(X,Y)* of the subject in the force-plate referential. X axis of the platform corresponds to the fore-aft direction and Y axis corresponds to the medio-lateral direction, whereas Z axis corresponds to the vertical axis (positive downwards).The rate frequency was 100 hertz for one duration of 60 seconds.

The subjects were placed on the force-plate in a cabin conformed to the standards [4] of the Association Française de Posturologie (AFP). They had to stand upright during one minute. The resting time between trials was one minutes too. They had to fix a plumb line placed at about 90 centimeters in front of them. The subjects' feet positions were controlled by a special frame fixed on the force-plate with a clearance of 2 cm between the feet and an angle of 15° between the medial sides of the feet.

*Recall on the method based on the stabilogram diffusion plot.*

The f.B.m. model is characterized by the scale law of the increments : the log/log diffusion plot as a function of increasing times scales $\Delta t$ is a straight line with a *2H* slope, where *H* is a real number *(0<H<1)* called the Hurst's parameter. Its meaning is of prime order : when $H$ is greater than 0.5 the increments are positively correlated and the process is persistent, when $H$ is less than 0.5 the increments are negatively correlated and the process is anti persistent.

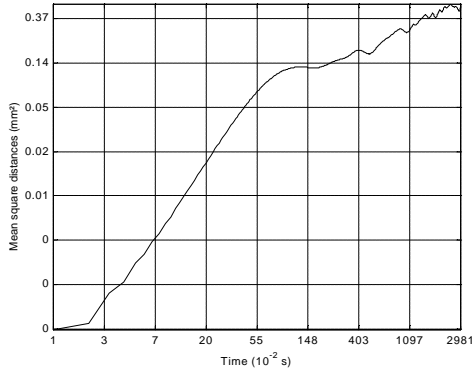

Fig. 1 The double logarithm plotting of the variogram of the X-axis COP which express the mean square distance $<\Delta X^2>$ as function of time scale $\Delta t$. The critical time scale is visually about $\Delta t_c = 1.05$ s. following [3] and $\Delta t_c = 0.67$ s. following [10].

The log/log plot of the diffusion corresponding to COP of human in upright quiet stance is not a straight line but exhibit at least two segments of straight line with a slope $2H_1$ with $H_1>0.5$ for small time scales and a slope $2 H_0$ with $H_0<0.5$ for large time scales [3, 6], see Fig. 1.

These two segments are interpreted as corresponding to two different behaviours : at small time scales the increments are persistent (this could not insure equilibrium) and at large time scales the increments are anti-persistent which insures the equilibrium. The critical point between these two segments is heuristically determined in [6] and these authors found a mean value of critical time scale $\Delta t_c = 0.9$ s. Another heuristic method is proposed in [6] which leads to a mean value of critical time scale $\Delta t_c = 0.4s$. and to a value of $H_1$ smaller than in [3]. In both studies [3, 6], the results are mostly empirical since the statistical treatment is not justified by any model.

*Brief recall on Fractional Brownian Motions*

Fractional Brownian motions were popularized after 1965 by Mandelbrot, see [5] for e.g. He first noticed the relevance of f.B.m. to describe and interpret various real phenomena and he developed the statistical study of f.B.m. during the following decade. However the f.B.m. were (implicitly) known before and were indeed introduced by Kolmogorov (1940). The representation of f.B.m. used by Mandelbrot in [5] (called Moving Average representation) supposes that the Hurst index $H$ is constant. In the model introduced by [3], there are different values depending on the frequency, anyway it is not possible to give any meaning to the Moving Average representation when the Hurst index $H$ is varying with the frequency. From the other hand, Kolmogorov use another representation called the harmonizable representation, which is a kind of Fourier representation. Moreover this representation is generalizable to every stationary Gaussian process and each process is caracterized by its density spectrum $\rho(f)$. F.B.m correspond to a power law density spectrum $r(f) = f^{-(H+\frac{1}{2})}$ where f is the frequency. When the Hurst index depend on the frequency, we denote it $H(f)$, this corresponds to the process with density spectrum $r(f) = f^{-(H(f)+\frac{1}{2})}$. It is then natural to consider the COP as a process with different Hurst indexes following the frequencies associated to the density spectrum $r(f) = f^{-(H(f)+\frac{1}{2})}$, with a function $H(f)$ piecewise constant. In [1], this model was called Multiscale fractional Brownian motion (m.s.f.B.m.). After having given a rigorous definition of the model introduced in [3], we want to estimate the value of the parameters of this model, *i.e.* the frequency changes and the different Hurst indexes.

Let us stress another famous property of f.B.m. and some consequence in our model. F.B.m. are self-similar processes, this means they have the same statistical behavior at every scales. More precisely if $B_H(t)$ is a f.B.m. with the Hurst index $H$, then $B_H(a\,t)$ as the same distribution than $a^H \cdot B_H(t)$. So f.B.m. corresponds to a stochastic fractal. The boundness of COP trajectories (which remain in the quadrilater defined by the two feet) induces that at very large scale the Hurst index should be 0 and that at least a second frequency change should exist.

*Brief recall on Wavelets Analysis*

Wavelets are a convenient tool to analyze the signal's properties at different scales. In many fields of signal processing, wavelet analysis has replaced the classical Fourier analysis. In Fourier analysis, a signal is decomposed onto a basis of functions $sin(\omega\,t)$ or $cos(\omega\,t)$. Each of these functions have only one frequency, but it have a infinite time support. Therefore Fourier analysis is convenient for stationary signal observed on a long time interval. Wavelets correspond to another tradeoff between time localization and frequency localization : wavelets take into account a finite band of frequencies and they are decreasing very fast in time (Fig. 2), this allows frequency analysis of a signal observed on a shorter time interval.

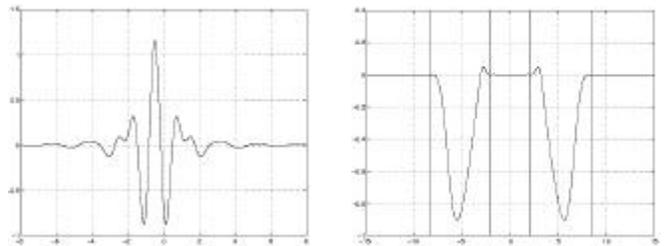

Fig. 2. a) the Meyer wavelet $\Psi(t)$
b) its Fourier transform $\hat{\Psi}(f)$

*Wavelet analysis of f.B.m. and m.s.f.B.m*

Then we analyze the signal through the wavelet coefficient at different scales *a* wich corresponds to the frequencies $f=1/a$. Let $\Psi(t)$ be the Meyer wavelet [8], the wavelet coefficient at frequency $f$ of the signal $X(t)$ observed with a time lag $\Delta$ is defined by

$$I(f) = \frac{1}{[Nf]}\sum_{k=1}^{[Nf]} d_X^2(f,k\Delta) \qquad (1)$$

where N is the number of points observed for the signal $X(t)$, , $\Delta$ the time lag between two observations, *[Nf]* the entire part of the real number *Nf* and

$$d_x(f,b) = \Delta f^{\frac{1}{2}} \sum_{p=0}^{N} \Psi(p\Delta f - b) X(p\Delta) \qquad (2)$$
$$\approx f^{\frac{1}{2}} \int \Psi(tf-b) X(t) dt$$

Formula (2) explains that the wavelet coefficient *I(f)* corresponds to the energy on the frequency band $[2\pi f/3, 8\pi f/3]$. When the time lag $\Delta$ is sufficiently small (which is equivalent to a sufficiently high frequency of sampling ), we have (see [1])

$$\log I_X(f) = \log \int_{2\pi/3}^{8\pi/3} \left|\hat{\Psi}(u)\right|^2 r^2(fu)\, du + \frac{1}{\sqrt{N\Delta}} e_f \qquad (3)$$

where $e_f$ is a centred Gaussian vector. If the signal $X(t)$ were a f.B.m. of Hurst index $H$, then the square density spectrum is $r^2(fu) = f^{-(2H+1)} \cdot u^{-(2H+1)}$ which implies

$$\log I_X(f) = -(2H+1)\log f + \log K_H(\Psi) + \frac{1}{\sqrt{N\Delta}} e_f \qquad (4)$$

where $K_H(\Psi) = \int_{2\pi/3}^{8\pi/3} \left|\hat{\Psi}(u)\right|^2 u^{-(2H+1)} du$. Therefore a linear regression of *log(I_X(f))* onto *log(f)* provides the slope $-(2H+1)$ and after the Hurst index *H*. This result remains in force as soon as the density spectrum satisfy the condition $r^2(u) = u^{-(2H+1)}$ with a constant Hurst index *H* when the frequency *u* belongs to the frequency band $[2\pi f/3, 8\pi f/3]$. When the C.O.P. follows the model of Collins & De Lucas, the coordinate $X(t)$ corresponds to a m.s.f.B.m. with one frequency change located in $\omega_1$ (in the frequency band studied) with a Hurst index $H_0$ for frequencies $f < \omega_1$ and with a Hurst index $H_1$ for frequencies $f > \omega_1$. In this case, we have a linear regression with slope $-(2H_0+1)$ when $f' = 4\pi f/3 < w_1/2$, a linear regression with slope slope $-(2H_1+1)$ when $f' = 2\pi f/3 > 2 \omega_1$ and a transition zone when $w_1/2 < f' < 2 \omega_1$, see Fig.3.

To avoid this transition zone, we compute the difference of the slope estimated by linear regression into left and right boxes of size $\Lambda$ separated by a hole of size $log\,(4) = log(2\,\omega_1) - log(\omega_1/2)$, see [1] and Fig. 4. The value of the critical frequency is estimated as the first minimum of the function *D(f)*.

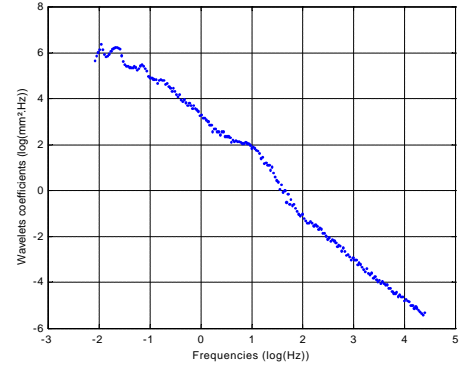

Fig.3 The double logarithm plotting of the wavelets coefficients of the X coordinate of the COP as a function of frequency (computed on the same data than the variogram of Fig. 1).

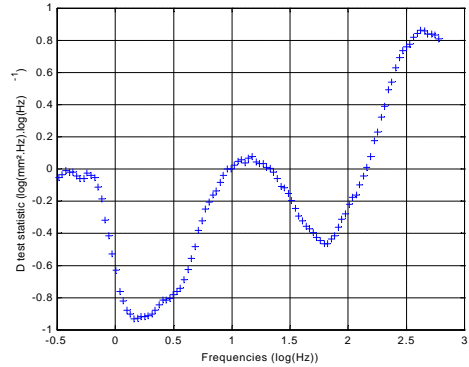

Fig. 4 The plot of the test statistic.
The function *D(f)* express the difference of the slope of Fig. 3 computed on two sliding boxes separates by a hole of size *log(4)*. The abscissa of the minimum of *D(f)* corresponds to the frequency change, in this case $log\,\omega_1 = 0.16$, $\omega_1 = 1.17\,Hz$ and $\Delta t_c = 0.67s$.

*Description of the method.*

i) We choose the algorithm parameters. We fix the frequency band $[\omega_{min}, \omega_{max}]$ in which we want to detect and estimate frequency changes on the Hurst index. We have chosen $\omega_{min} = 0.633\,Hz$ and $\omega_{max} = 16.62\,Hz$ which correspond to time scales *1.57 s.* and *0.06 s*. We fix the discretization on frequencies $q = 4^{1/45}$ and the number of frequencies in the left and the right box $\Lambda = 30$. Those parameters automatically determine the number of frequencies *M*, here *M= 211* and the ratio $r = 2.q^\Lambda = 5.04$.

ii) We compute the logarithm of the wavelets coefficient *log I(f_k)* for frequencies $f_k = q^k \cdot \omega_{min}/r$ for *k =1,...,M* , (Fig. 3).

iii) We compute $D(f_k)$ defined as the difference between the linear regression coefficients (in log/log scale) within a left box (for frequencies $f_j$, with an index *j* in *[k-22-**L**, k-23]*) and within a right box (for frequencies $f_j$, with an index *j* in *[k+22, k+23+$\Lambda$]*). The left and right boxes are separated by a hole of 45 indexes which corresponds to a size *log 4 = 45 log q*, this is the length of the transition zone

iii) We determine the critical frequency $\omega_1$ as the first minimum of the function *D(f)* and after the two Hurst indexes

$H_0$ (resp. $H_1$) from the slopes of the segments of Fig. 3 with abscissa $f$ in $[\omega_{min}, \omega_1/r]$, resp. $f$ in $[r.\omega_1, \omega_{max}]$.

The program is written with Matlab 5.2 and uses the wavelets toolbox which provides the values of the Meyer wavelet $\Psi(t)$ and of its Fourier transform. Fig. 3 gives an example and shows the frequency change at *log f =0.16*, that is *f =2.38 Hz*.

### III. RESULTS

This method has been applied our method to a set of 16 subjects and the results are compared with the variogram method proposed in [10].

TABLE I

|  | Log freq. | Freq. (Hz) | Time (s.) | Time[10] (s.) |
|---|---|---|---|---|
| Mean | 0.48 | 1.80 | 0.68 | 0.53 |
| (Std) | (0.47) | (0.93) | (0.28) | (0.26) |

Critical frequencies vary from *1.04 Hz* to *4.14 Hz* corresponding to critical time scales from *0.24 s* to *0.96 s*. For the other hand, the method proposed in [10] leads on the same set of data to time scales varying from *0.02s* to *1.01s*. For the data corresponding to Fig.1, we have $\Delta t_c = 0.85$ s when [10] gives $\Delta t_c = 0.67$ s. We find, as in [3] and [10] a Hurst index $H_0<0.5$ for small frequencies corresponding to large time scales and a Hurst index $H_1>0.5$ for large frequency or small time scales.

### IV. DISCUSSION

The wavelet analysis of the COP signal leads most often to critical time scales between the critical time scale provided by [10] and the critical time scale provided by [3] and close to the second one. But the method in [3] is effectively visual and heuristic, it depends on the operator. Moreover the critical time scales could depend on the duration of the observation : we obtain different critical time scales when the duration is 20 seconds, 40 seconds or 60 seconds, see [1], this is due to the existence of at least another frequency change on the Hurst coefficient at large scales. The frequency localization of Fourier transform of the Meyer wavelet induces a propagation of a frequency change only on a finite band of frequency of size *log 4* in *log Hz*. For this reason the wavelet algorithm is robust with respect to the duration of the observation. Last but not least, the proposed wavelet analysis is based onto a mathematical model well constructed which protects us against numerical artifacts.

From the biological point of view, the precise determination of the critical frequency is very relevant, more than the exact value of the Hurst index corresponding to the different control mechanism. The only relevant fact is that $H_0<0.5$ for small frequencies $H_1>0.5$ for large frequency. Precise knowledge of the critical frequency $\omega_1$ in given experimental conditions induces that only physiological phenomenon of frequency $f$ greater than $\omega_1$ contribute to the task of maintaining upright position. Making varying the experimental condition could then give new information about the control of equilibrium of human been.

### V. CONCLUSION

This algorithm is robust and allows an automatic detection of the critical frequency or the associated critical time scale. Operators subjectivity is so avoid.

With this tool, it is then possible to describe the different factors effects on the postural control mechanism : vision, feet position, feedback, gender. Other challenging way is to explore the relationship between critical frequency (or critical time scale) and physiological phenomenon.

### REFERENCES

[1] P.Bertrand and J.M. Bardet, "Some generalization of fractional Brownian motion and Control", in *Optimal Control and P.D.E.*, J.L. Menaldi, E. Rofman and A. Sulem Eds., IOS Press, 2001, pp.221-230.

[2] P. Bertrand, "A local method for estimating change points : the hat-function", *Statistics* 34, 2000, pp. 215-235.

[3] J.J. Collins and C.J., De Luca, "Open-loop and closed-loop control of posture : A random walk analysis of center-of-pressure trajectories", 1993, *Exp. Brain. Res.* **95**, pp.308-318

[4] P. M. Gagey, B. Asselain, N. Ushio and J. B. Baron, *Huit leçons de posturologie* 4$^{th}$ ed., vol 4 Paris : Association Française de Posturologie, 1994, pp.2-4.

[5] P.M. Gagey and B. Weber, *Posturologie : régulation et dérèglement de la station debout*, Paris, Masson, 1995.

[6] A.N. Kolmogorov, "Wienersche Spiralen und einige andere interessante Kurven in Hilbertschen Raum", *Doklady* **26**, 1940, pp. 115-118.

[7] B.B. Mandelbrot and J.W. Van Ness, "Fractional Brownian motion, fractional noises and applications", *SIAM review* **10**, 1968, pp.422-437.

[8] Y. Meyer, *Wavelets: Algorithms and Applications*, SIAM, 1993.

[9] K.M. Newell, S.M. Slobounov, E.S. Slobounova and P.C.M. Molenaar, "Stochastic processes in postural center-of-pressure profiles", *Exp. Brain Res.* **113**, 1997, pp.158-184.

[10] P. Rougier, "Influence of visual feedback on successive control mechanisms in upright quiet tance in humans assessed by fractional Brownian motion modelling", *Neuroscience Letters*, 266, 1999, pp.157-160.